\documentclass[aps,prl,twocolumn,showpacs,superscriptaddress,groupedaddress]{revtex4}  
\usepackage{amssymb}   
\usepackage{amsmath}
\usepackage{graphicx}
\usepackage{subfig}
\usepackage{color}

\def\beq{\begin{equation}}
\def\eeq{\end{equation}}
\def\bsp#1\esp{\begin{split}#1\end{split}}

\newcommand\refr[1]     {ref.\,\cite{#1}}

\newcommand\eqn[1]     {eq.\,(\ref{#1})}

\newcommand\fig[1]     {fig.~{\ref{#1}}}
\newcommand\figs[2]    {figs.\,\ref{#1} and~\ref{#2}}

\def\beq{\begin{equation}}
\def\eeq{\end{equation}}
\def\bsp#1\esp{\begin{split}#1\end{split}}
\def\bal#1\eal{\begin{align}#1\end{align}}

\newcommand\tsig[2]    {\sigma^{\rm{#1}}_{#2}}
\newcommand\dsig[2]    {\rd\sigma^{{\rm #1}}_{#2}}
\newcommand\dsiga[3]   {\rd\sigma^{{\rm #1,A}_{\scriptscriptstyle #2}}_{#3}}

\newcommand{\rd}       {{\rm{d}}}

\newcommand{\cSCS}[2]  {{\cal C}\kern-2pt{\cal S}_{#1}^{#2}}

\newcommand{\rC}       {{\rm C}}
\newcommand{\rS}       {{\rm S}}
\newcommand{\rSCS}     {{\rC}\kern-2pt{\rS}}

\newcommand{\IcSCS}[2] {\rC\kern-2pt\rS_{#1}^{#2}}

\newcommand{\colorfulNNLO}{{CoLoRFulNNLO}}

\begin{document}

\begin{flushleft} 
\mbox{CERN-PH-TH/2016-066, CP3-16-09}
\end{flushleft}

\title{Three-jet production in electron-positron collisions using the \colorfulNNLO\ method}

\author{Vittorio Del Duca}
\altaffiliation{On leave from INFN, Laboratori Nazionali di Frascati, Italy.}
\affiliation{Institute for Theoretical Physics, ETH Z\"urich, 8093 Z\"urich, Switzerland}

\author{Claude Duhr}
\altaffiliation{On leave from the ``Fonds National de la Recherche Scientifique'' (FNRS), Belgium.}
\affiliation{CERN Theory Division, 1211 Geneva 23, Switzerland}
\affiliation{Center for Cosmology, Particle Physics and Phenomenology (CP3),
Universit\'{e} Catholique de Louvain, \\
1348 Louvain-La-Neuve,
  Belgium} 

\author{Adam Kardos}
\affiliation{University of Debrecen and MTA-DE Particle Physics Research Group, H-4010 Debrecen, PO Box 105, Hungary}

\author{G\'abor Somogyi}
\affiliation{University of Debrecen and MTA-DE Particle Physics Research Group, H-4010 Debrecen, PO Box 105, Hungary}

\author{Zolt\'an Tr\'ocs\'anyi}
\affiliation{University of Debrecen and MTA-DE Particle Physics Research Group, H-4010 Debrecen, PO Box 105, Hungary}

\date{\today}

\begin{abstract}
We introduce a subtraction method for jet cross sections at
next-to-next-to-leading order (NNLO) accuracy in the strong coupling
and use it to compute event shapes in three-jet production in
electron-positron collisions.  We validate our method on two event
shapes, thrust and $C$-parameter, which are already known in the
literature at NNLO accuracy and compute for the first time oblateness
and the energy-energy correlation at the same accuracy.
\end{abstract}

\maketitle

One of the most important fundamental parameters in the standard model
is the strong coupling $\alpha_s$.  A clean environment for determining
$\alpha_s$ is the study of event shape distributions in $e^+\,e^-$
collisions~\cite{Kunszt:1989km}. Indeed, while at leading order (LO)
the production of two jets is a purely electroweak process, the
dominant contribution to the production rate of every additional jet in
the final state is directly proportional to the strong coupling.  Since
the initial state does not involve colored partons, non-perturbative
QCD corrections are restricted to hadronization and power corrections
affecting the final state configuration.  These corrections can be
determined either by extracting them from data by comparison to Monte
Carlo predictions or by using analytic models. The precision of the
theoretical
predictions is thus mostly limited by the accuracy in the perturbative
expansion in the strong coupling. Currently, the state of the art
includes next-to-leading order (NLO) predictions for the production of
up to five jets~\cite{Signer:1996bf,Frederix:2010ne} (and up to seven
jets in the leading color approximation~\cite{Becker:2011vg}), and
next-to-next-to-leading order (NNLO) predictions for the production of
three jets~\cite{Gehrmann-DeRidder:2007nzq,Weinzierl:2008iv}. 
Moreover, fixed-order predictions can be matched to resummation
calculations (see examples in ref.~\cite{Catani:1992ua}), which take
into account classes of logarithmically-enhanced contributions to all
orders in perturbation theory.  

The goal of this letter is twofold. First we present a framework to
compute fully differential predictions at NNLO accuracy for processes
with a colorless initial state and involving any number of colored
massless particles in the final state.  Second we apply our method
to event shape observables with at least three hard final-state partons. 
As our framework allows for a fully differential
description of the final state, it puts no restriction beyond infrared safety 
on the class of observables that we can consider. The cornerstone of our framework 
is the \colorfulNNLO\ (Completely Local subtRactions for Fully differential 
predictions at NNLO accuracy) method to regularize infrared
divergences~\cite{Somogyi:2005xz}. The method is based on the universal
factorization of QCD matrix elements in soft and collinear limits.
It takes into account all spin and color-correlations among the
final-state particles and as a result the subtractions are completely local.

Predictions at NNLO in perturbative QCD generically require the
computation of two-loop corrections to the Born process, as well as
one-loop and tree-level contributions to the processes with one or two
additional partons in the final state. The two-loop matrix elements for
$\gamma^*/Z\to q\,\bar{q}\,g$ have been computed in ref.~\cite{Garland:2001tf}.
The one-loop corrections to four-jet production have been computed in
ref.~\cite{Glover:1996eh}, and the tree-level matrix elements for the production of
five jets were obtained for first time in ref.~\cite{Hagiwara:1988pp}. 
The sum of these contributions is finite for infrared-safe observables, 
but taken separately, they all exhibit explicit divergences coming from loop 
integrations and/or implicit divergences when one or more partons in the final 
state are unresolved. Thus each contribution needs to be separately rendered finite 
in four dimensions before any numerical computation can be performed.  

The \colorfulNNLO\ method deals with this issue by using universal
counterterms to redistribute divergences between different final-state
multiplicities. In every singular region of phase space we subtract the
corresponding infrared divergence through a suitably constructed
approximate matrix element. As different singular regions overlap, a
careful bookkeeping is required in order to avoid double-counting when
subtracting the divergences. Moreover, within our framework beyond NLO,
we also need to consider iterated singular limits where partons become
successively unresolved. Finally, spin correlations in gluon decay are 
retained, and the counterterms are fully local in phase space. 
In the \colorfulNNLO\ framework the distribution of the NNLO correction
to an observable $J$ can be written as a sum of three contributions, 
each being separately finite in $d=4$ dimensions,
\beq 
\tsig{NNLO}{}[J] = \!\!
	\int_{m+2}\!\!\!\!\!\dsig{NNLO}{m+2} 
	+ \int_{m+1}\!\!\!\!\!\dsig{NNLO}{m+1} 
	+ \int_m\!\dsig{NNLO}{m}\,,
\label{eq:sigmaNNLOfin} 
\eeq 
where
\bal
\label{eq:sigmaNNLOm+2} 
\dsig{NNLO}{m+2} &= 
	\Big\{\dsig{RR}{m+2} J_{m+2} 
	- \dsiga{RR}{2}{m+2} J_{m} \\
\nonumber&	-\Big[\dsiga{RR}{1}{m+2} J_{m+1} 
	- \dsiga{RR}{12}{m+2} J_{m}\Big]\Big\}_{d=4}\,, 
\\ 
\label{eq:sigmaNNLOm+1} 
\dsig{NNLO}{m+1} &= 
	\Big\{\Big[\dsig{RV}{m+1} 
	+ \int_1\dsiga{RR}{1}{m+2}\Big] J_{m+1}  \\
\nonumber&	-\Big[\dsiga{RV}{1}{m+1} 
	+ \Big(\int_1\dsiga{RR}{1}{m+2}\Big)\strut^{{\rm A}_{\scriptscriptstyle 1}} 
	\Big] J_{m} \Big\}_{d=4}\,,
\\
\label{eq:sigmaNNLOm}
\dsig{NNLO}{m} &= 
	\Big\{\dsig{VV}{m} 
	+ \int_2\Big[\dsiga{RR}{2}{m+2} 
	- \dsiga{RR}{12}{m+2}\Big] \\
\nonumber&	+\int_1\Big[\dsiga{RV}{1}{m+1} 
	+ \Big(\int_1\dsiga{RR}{1}{m+2}\Big) \strut^{{\rm A}_{\scriptscriptstyle 1}} 
	\Big]\Big\}_{d=4} J_{m}\,.
\eal
$J_n$ denotes the value of the infrared-safe observable $J$
evaluated on a final state with $n$ resolved partons.  The various
subtraction terms have been defined explicitly and their integrals over
the factorized phase space of the unresolved partons have been obtained
in \refr{Somogyi:2005xz}.

Eq.~\eqref{eq:sigmaNNLOm+2} includes the double-real (RR)
contribution that exhibits singularities whenever one or two partons
become unresolved. We subtract from this an approximate cross section
$\dsiga{RR}{2}{m+2}$ which has the same singularities in the
doubly-unresolved limits as the RR matrix element. The difference
is still singular in singly-unresolved regions and an
additional counterterm is needed. To that effect, we subtract the
quantity $\dsiga{RR}{1}{m+2}$, and compensate for the overlap between
the singly and doubly-unresolved regions through the term
$\dsiga{RR}{12}{m+2}$.  

Similarly, eq.~\eqref{eq:sigmaNNLOm+1} describes the emission of one
additional parton at one loop, the real-virtual (RV) contribution. In
addition to explicit infrared poles coming from the loop integration,
the RV contribution has kinematic singularities when this additional
parton is soft or collinear to another colored particle. These are
regularized by the approximate one-loop cross section
$\dsiga{RV}{1}{m+1}$. The difference is now free of implicit
singularities, but the explicit infrared singularities are still
present. These poles are the same as the singly-unresolved singularities
in $\dsiga{RR}{1}{m+2}$ in integrated form, and they cancel once the
corresponding term is added back. The integral of $\dsiga{RR}{1}{m+2}$,
however, still has one-parton singularities, which are regularized by
the last term in eq.~\eqref{eq:sigmaNNLOm+1}.  

The last contribution to the NNLO distribution, shown in
eq.~\eqref{eq:sigmaNNLOm}, includes two-loop virtual (VV) corrections
to the LO process. The two-loop integrations lead to explicit infrared
singularities. In eqs.~\eqref{eq:sigmaNNLOm+2} and~\eqref{eq:sigmaNNLOm+1} 
we had introduced five counterterms, but so far only $\dsiga{RR}{1}{m+2}$ 
has been added back in integrated form. The explicit two-loop singularities 
cancel against the phase-space singularities of the remaining four
counterterms, which are shown explicitly in integrated forms in
eq.~\eqref{eq:sigmaNNLOm}.  We computed all terms in these integrated
forms that become singular in $d=4$ dimensions and demonstrated the
cancellation of these divergences analytically. We also computed
analytically the logarithmic terms in the finite part of the integrated
subtractions that become singular on the edges of the phase space,
while evaluated the rest of the finite part of the integrated
subtractions numerically. We add the uncertainty of these numerical
evaluations to the uncertainty of the Monte Carlo integration of the
$n$-parton integrations in eq.~\eqref{eq:sigmaNNLOfin} in quadrature.  

\colorfulNNLO\ has already been successfully applied to compute NNLO
corrections to differential distributions describing the decay of a Higgs boson
into a pair of $b$-quarks~\cite{DelDuca:2015zqa}.  Here we apply this framework for
the first time to the computation of NNLO observables with more than
two colored partons in the final state. In particular, we consider
event-shape observables in $e^+\,e^-\to \gamma^* \to 3$ jets and we study NNLO
corrections to them.  If $\mathcal{O}$ denotes a generic event shape
observable,
we write 
\beq
\bsp
\frac{1}{\sigma_0}
\frac{\rd \sigma}{\rd \mathcal{O}} &=
  \frac{\alpha_s}{2\pi}A(\mathcal{O})
+ \left(\frac{\alpha_s}{2\pi}\right)^2 B(\mathcal{O})
+ \left(\frac{\alpha_s}{2\pi}\right)^3 C(\mathcal{O})
\\
&+ \mathcal{O}(\alpha_s^4)
\,,
\label{eq:Oexpansion}
\esp
\eeq
where $\sigma_0$ is the leading-order prediction for the process
$e^+\, e^- \to$ hadrons in perturbation theory.  In this letter we
concentrate on four event shapes. The first two,
thrust~\cite{Brandt:1964sa,Farhi:1977sg} and 
$C$-parameter\,\cite{Parisi:1978eg,Donoghue:1979vi}, have already been
studied at NNLO accuracy~\cite{Gehrmann-DeRidder:2007nzq,Weinzierl:2008iv}
and serve as a validation of our method. The other two, oblateness
and energy-energy correlation, have never been presented at NNLO
accuracy and constitute our main phenomenological results. 
Thrust is defined as
\beq
T = \max_{\vec n} \left( \frac{\sum_i |\vec n\cdot \vec p_i|}{\sum_i |\vec p_i|} \right) \,,
\label{eq:thrust}
\eeq
where $\vec p_i$ denote the three-momenta of the partons and $\vec n$
defines the direction of the thrust axis, $\vec n_T$, by maximizing the
sum on the right-hand side over all directions of the final-state
particles. In order to define oblateness, we need two variants of this
definition, thrust major and thrust minor. Thrust major is given by
\beq
T_M = \max_{\vec n\cdot \vec n_{T} = 0} \left( \frac{\sum_i |\vec n\cdot \vec p_i|}{\sum_i |\vec p_i|} \right) \,,
\label{eq:thrustmaj}
\eeq
where $\vec n$ defines the direction of the thrust-major axis,
$\vec n_{T_M}$, by maximizing the sum on the right-hand side over all
directions orthogonal to the thrust axis.  Similarly,
\beq
T_m = \frac{\sum_i |\vec n_{T_m}\cdot \vec p_i|}{\sum_i |\vec p_i|} \,, \quad {\rm with} \quad 
\vec n_{T_m} = \vec n_T \times \vec n_{T_M}\,,
\label{eq:thrustmin}
\eeq
defines thrust minor,
where the thrust-minor axis, $\vec n_{T_m}$, is orthogonal
to both the thrust and thrust-major axes. Oblateness $O$ is then 
the difference of thrust major and thrust minor~\cite{Barber:1979yr}, 
\beq 
O = T_M - T_m\,.
\eeq

The value of the $C$-parameter for massless final-state particles is
\beq
\bsp
C_{\rm par} &=
 \frac3{2} \frac{\sum_{i,j} |\vec p_i| |\vec p_j| \sin^2\theta_{ij}}{\left(\sum_i |\vec p_i| \right)^2}\,,
\label{eq:Cpar}
\esp
\eeq
where $\theta_{ij}$ is the angle between $\vec p_i$ and $\vec p_j$. 

Finally, energy-energy correlation~\cite{Basham:1978bw} is the
normalised energy-weighted cross section defined in terms of the angle
between two particles $i$ and $j$ in an event,
\beq\bsp
{\rm EEC}(\chi) &\,= \frac1{\sigma_{\rm had}} \sum_{i,j} \int \frac{E_i E_j}{Q^2} \\
&\qquad\times\rd \sigma_{e^+e^-\to i\, j + X}
\delta(\cos\chi+\cos\theta_{ij})\,,
\label{eq:eec}
\esp\eeq
where $Q^2$ is the squared center-of-mass energy, $E_i$ and $E_j$ are
the particle energies, $\theta_{ij} = \pi - \chi$ is the angle between
the two particles and $\sigma_{\rm had}$ is the total hadronic cross
section. Experience shows that computing radiative corrections to the
distributions of $C$-parameter, oblateness and energy-energy correlations
is numerically more challenging than for other three-jet event shapes.

\begin{figure}[!t]
\includegraphics[scale=0.5]{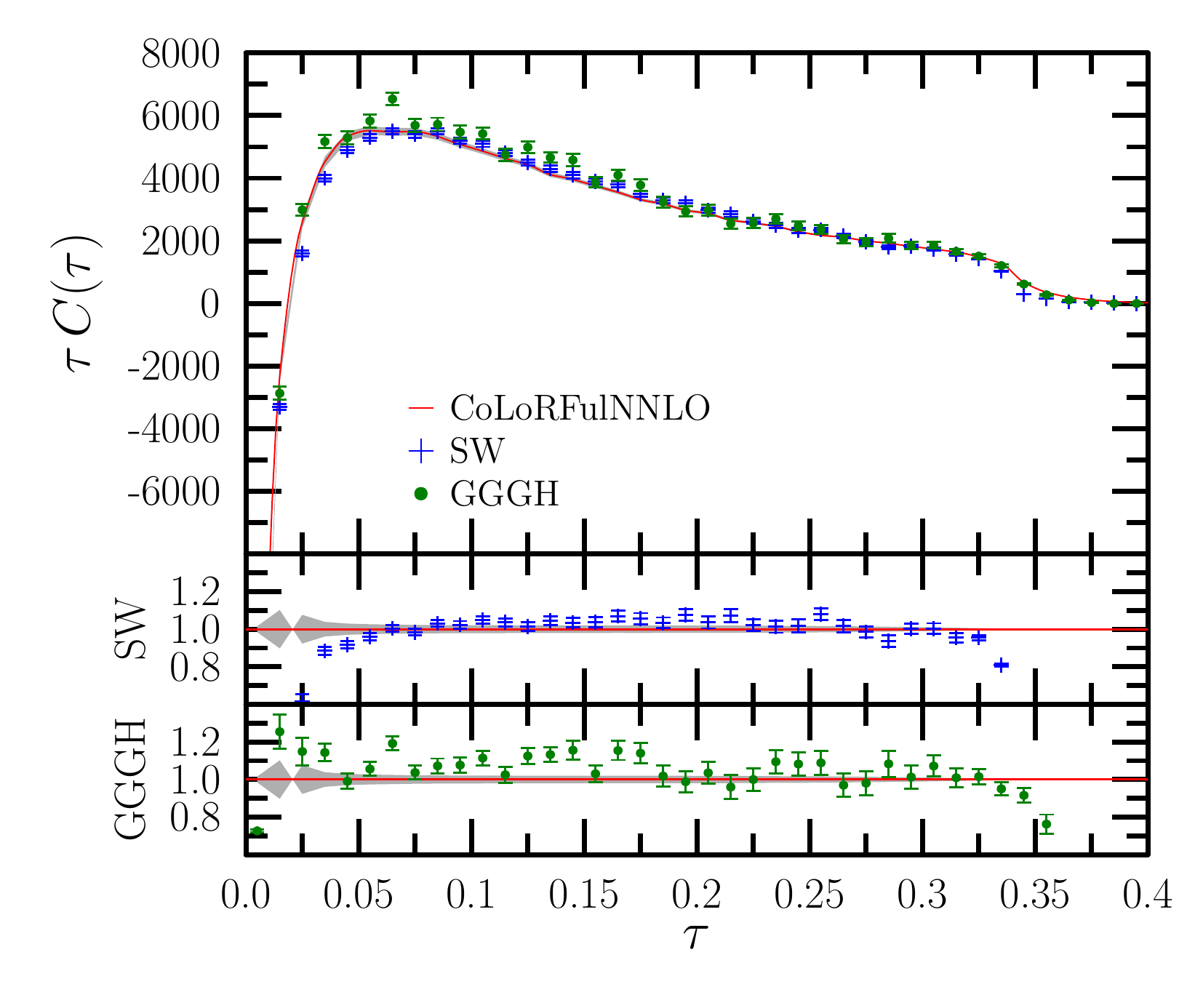}
\caption{\label{fig:thrust}The NNLO coefficient of the weighted
$\tau = 1-T$ distribution. The lower panels show the predictions of
ref.~\cite{Weinzierl:2008iv}, denoted as SW, (middle panel)
and those of ref.~\cite{Gehrmann-DeRidder:2007nzq}, denoted as GGGH,
(lower panel) normalized to ours, as well as the relative 
uncertainties of the numerical integrations
(shaded band around the line at one).
}
\end{figure}

\begin{figure}[!t]
\includegraphics[scale=0.5]{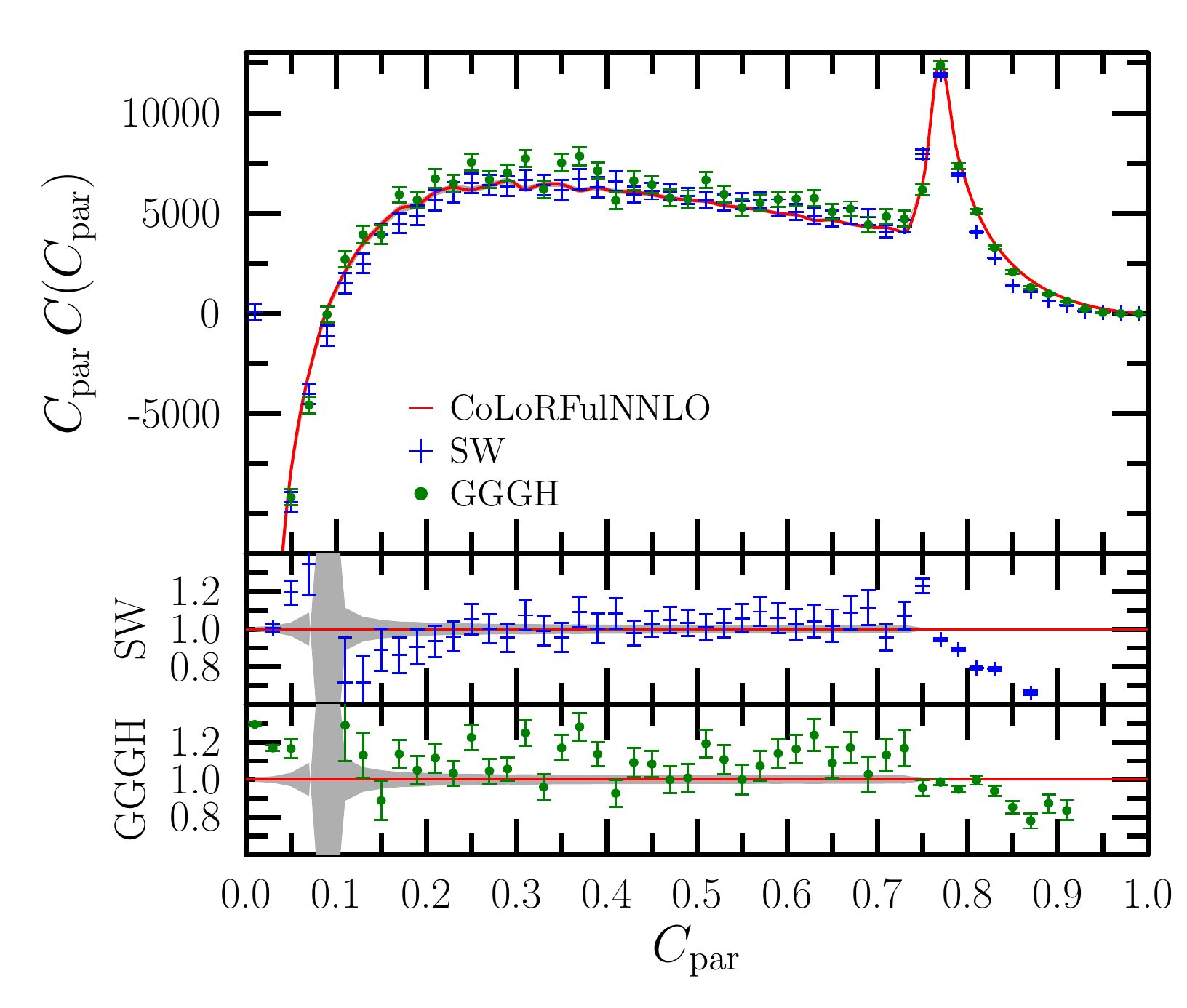}
\caption{\label{fig:C}The same as \fig{fig:thrust} for the
$C$-parameter.}
\end{figure}
As a validation of our method, we show in \figs{fig:thrust}{fig:C}
the third-order coefficient in \eqn{eq:Oexpansion} for
$\mathcal{O} = \tau \equiv 1-T$ and $\mathcal{O} = C_{\rm par}$. 
We observe a very good numerical convergence of our method at NNLO: the
absolute uncertainties of the integrations are shown as shaded narrow
bands around the solid line on the upper panels (hardly visible) and the
relative ones around the lines at one on the lower panels of
\figs{fig:thrust}{fig:C}.  We compare our results to the
predictions of refs.~\cite{Gehrmann-DeRidder:2007nzq,Weinzierl:2008iv}
and we find agreement over a large range of $\tau$ and $C$-parameter.
We observe statistically significant differences beyond the kinematical
limits ($\tau = 1/3$ and $C_{\rm par}=3/4$) where the three-particle
final states vanish and the event shapes are determined by a four-jet
final state.  In these regions the $C(\mathcal{O})$ coefficients are
determined by the NLO corrections to four-jet production, which have
been known for long \cite{Signer:1996bf} and can also be computed with
modern automated tools, such as {\tt MadGraph5\_aMC@NLO}
\cite{Alwall:2014hca}. We  have checked that our predictions are in
complete agreement with those of {\tt MadGraph5\_aMC@NLO}.

\begin{figure}[!t]
\includegraphics[scale=0.5]{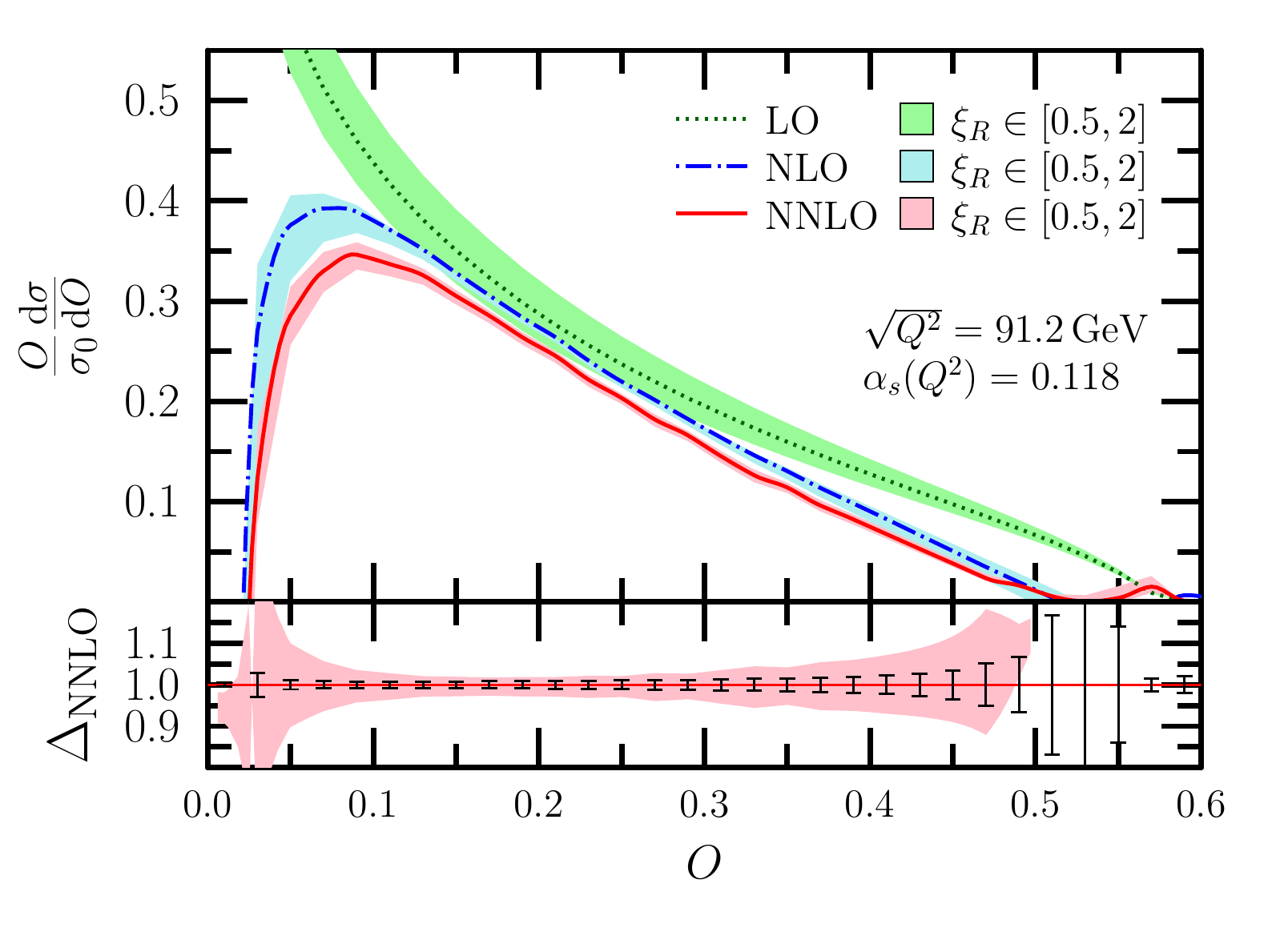}
\caption{\label{fig:Obl}Weighted distributions of oblateness $O$
at LO, NLO and NNLO accuracy in perturbative QCD. The bands represent
the dependence on the renormalization scale varied in the range 
$\xi_R\equiv\mu/\mu_0 \in [0.5, 2]$ around the default scale $\mu_0 =
\sqrt{Q^2}$.  The lower panel shows the relative scale dependence
(band) at NNLO accuracy and the relative uncertainty of the
integrations (error bars).
}
\end{figure}

We present predictions for the distributions of oblateness $O$ and
energy-energy correlation EEC at NNLO accuracy in perturbative QCD for
collider energy $\sqrt{Q^2}=91.2$\,GeV in \figs{fig:Obl}{fig:EEC}.  The
bands represent the dependence of the predictions on the
renormalization scale varied in the range $[0.5,2]$ times our default
scale: the total center-of-mass energy.  We use $\alpha_s = 0.118$ for
the central value and the three-loop running of the strong coupling for
computing the scale variations. The lower panels show the relative
scale dependence of the NNLO predictions and the relative uncertainties
of the integrations. Both oblateness and energy-energy correlation are
known to vanish in the dijet limit.  Moreover, oblateness is expected
to vanish also for cylindrically symmetric final states, while for
three-parton events one has $0\le O\le 1/\sqrt{3}$. Indications of
these features are visible in \figs{fig:Obl}{fig:EEC}.
 
\begin{figure}[!t]
\includegraphics[scale=0.5]{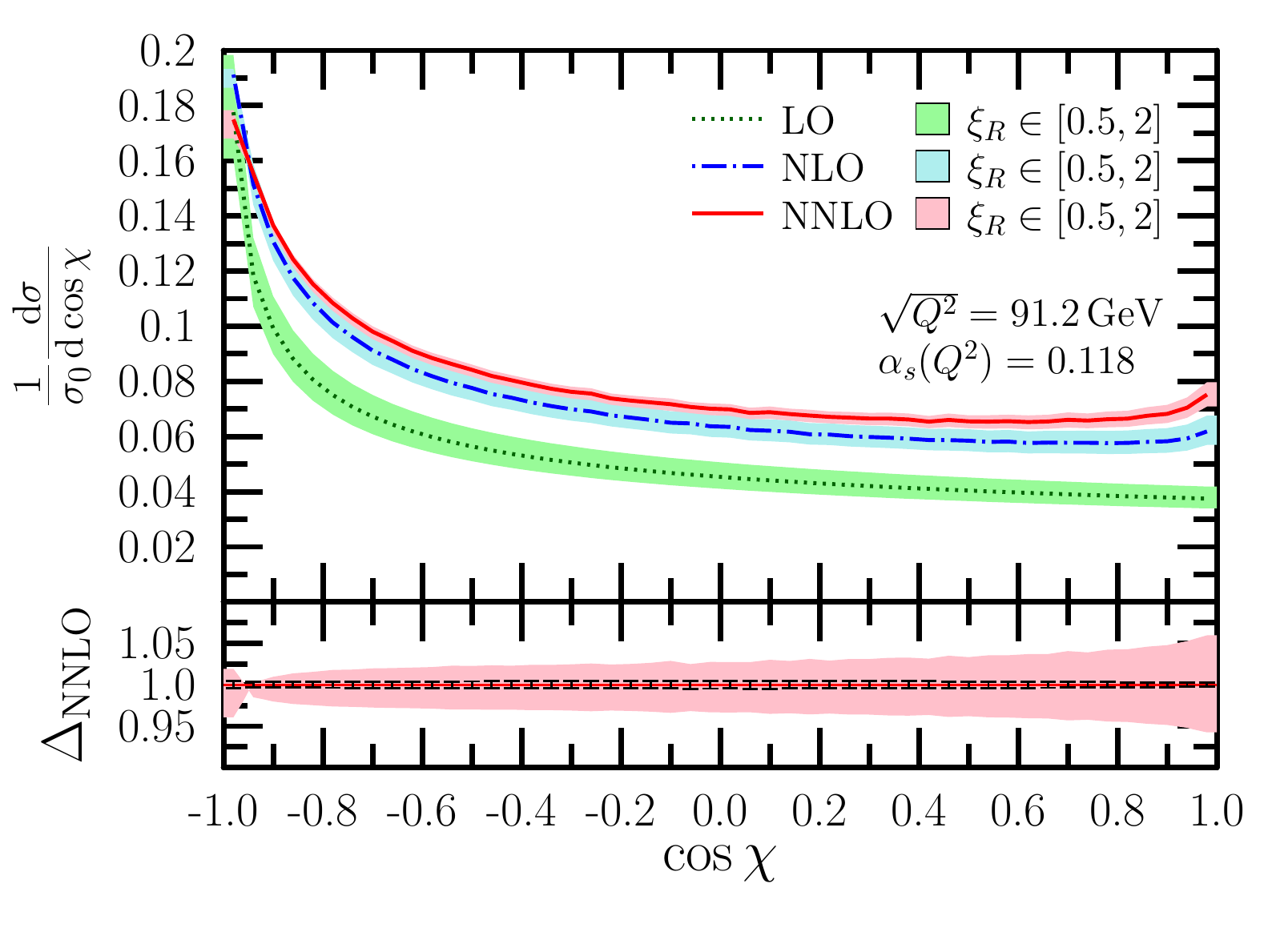}
\caption{\label{fig:EEC}Distributions of 
energy-energy correlation EEC at LO, NLO and NNLO accuracy in
perturbative QCD. The bands and the lower panel are like in \fig{fig:Obl}.}
\end{figure}

\begin{figure}[!t]
\includegraphics[scale=0.5]{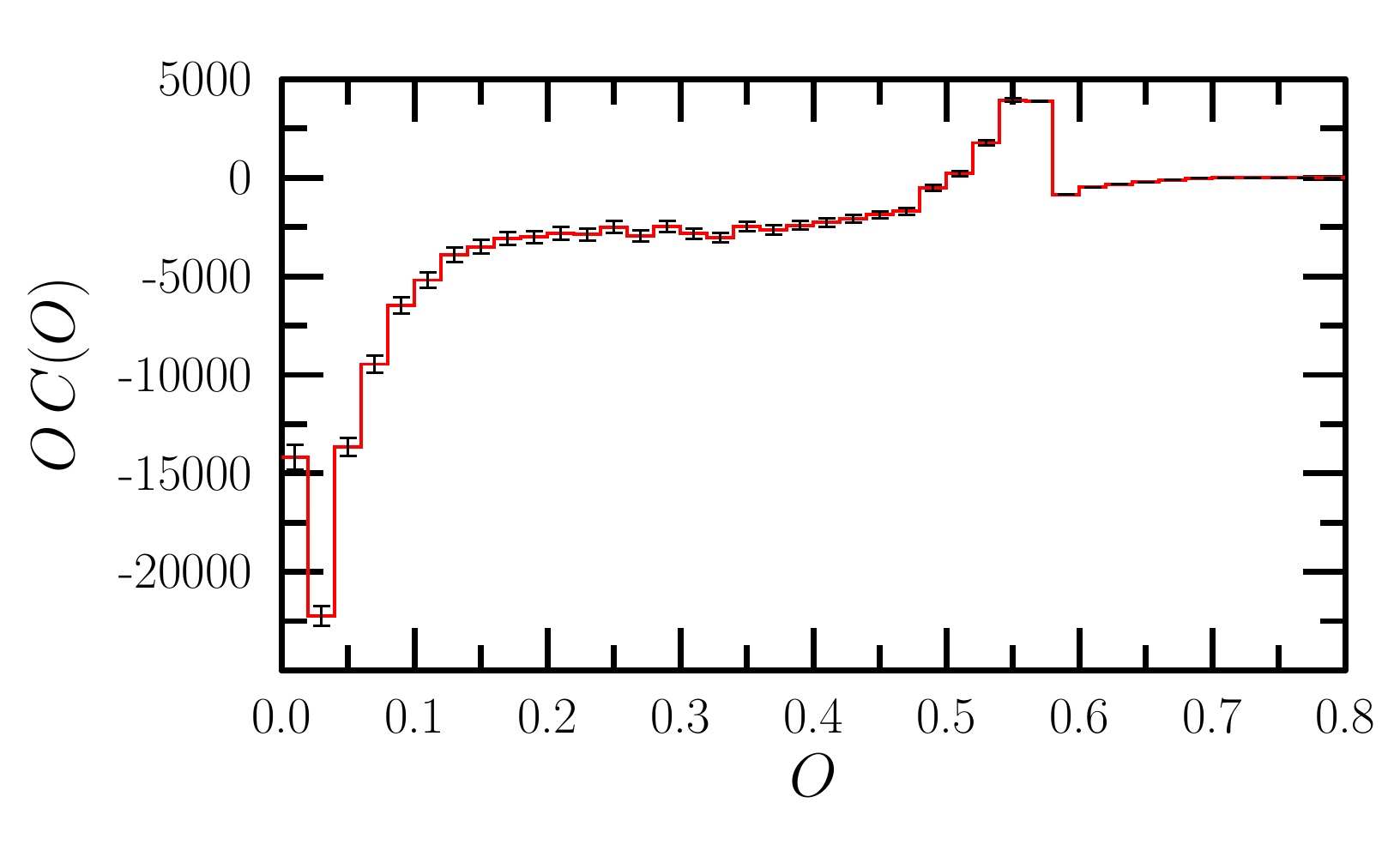}
\caption{\label{fig:Obl-C}Distribution of the NNLO coefficient
for oblateness $O$. The error bars represent the statistical
uncertainty of the Monte Carlo integrations.}
\end{figure}

We observe that the NNLO corrections slightly lower, and also
slightly modify the shape of the $O$-distribution compared to NLO,
while the NNLO corrections enhance the EEC-distribution almost uniformly. 
The changes in the shapes of the distributions due to the
NNLO corrections can be appreciated better by looking at the
distributions of the NNLO coefficients directly, as shown in
\figs{fig:Obl-C}{fig:EEC-C}.  Also for these
distributions, we observe good numerical convergence of our code.

\begin{figure}[!t]
\includegraphics[scale=0.5]{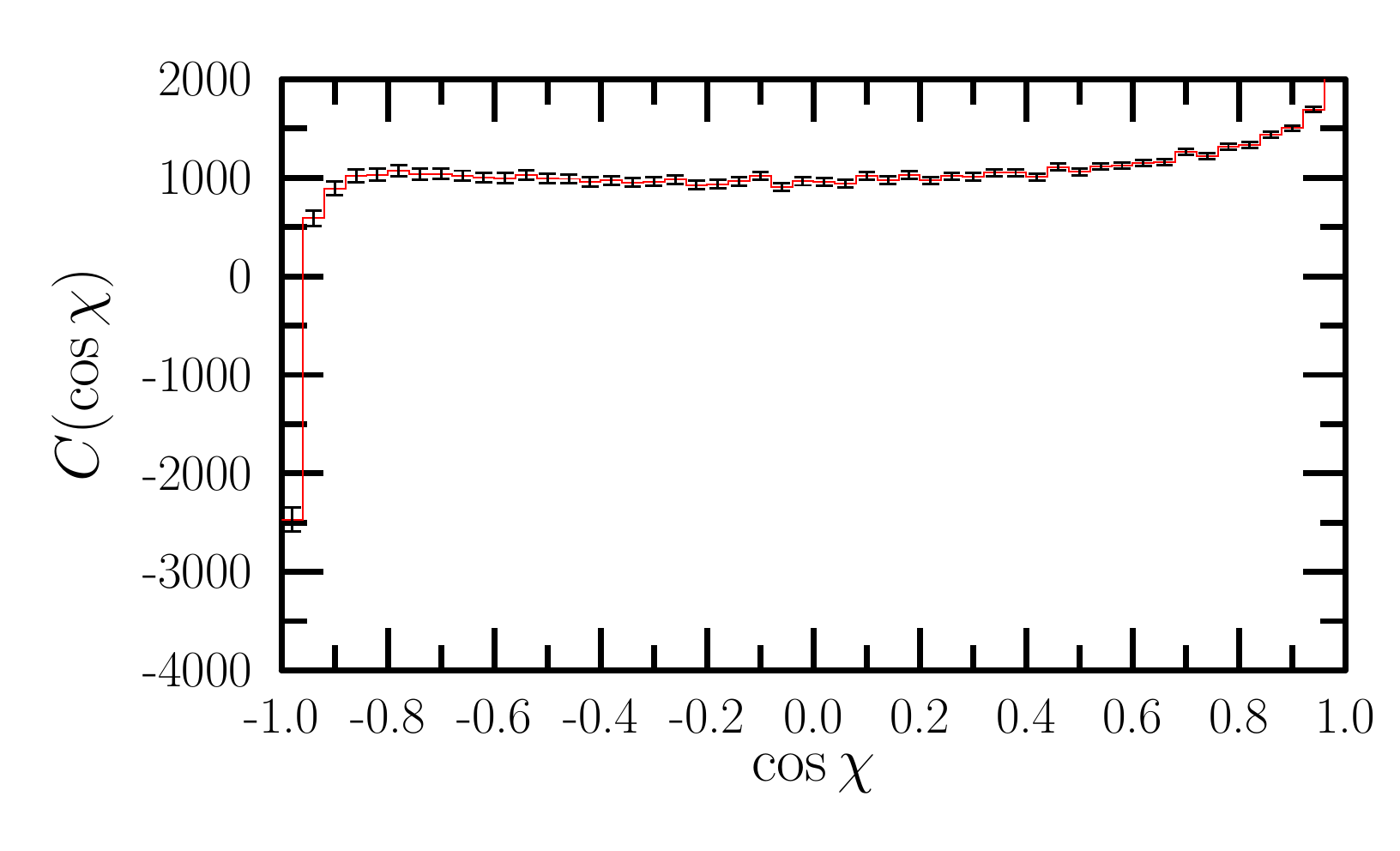}
\caption{\label{fig:EEC-C} Same as \fig{fig:Obl-C} for
energy-energy correlation.}
\end{figure}

We conclude by commenting on the behavior of the distributions
corresponding to small values of the event shapes. Those regions are
dominated by kinematical configurations where one of the three partons
is unresolved, resulting in logarithmically enhanced contributions.  In
order to obtain reliable predictions the large logarithms must be
resummed to all orders in perturbation theory, which is beyond the
scope of the present study.

In this letter we have introduced the \colorfulNNLO\ method to compute
NNLO radiative corrections for processes with colorless initial states.
We have applied it to obtain precise predictions for event shape
distributions in three-jet production in $e^+\,e^-$ collisions. We
observe very good numerical convergence of our predictions over the
whole range of values of the event shapes.  We emphasize that our
framework is not restricted to three-jet production, but it can be
easily extended to study differential distributions for four- or more
jet production once the corresponding two-loop amplitudes become
available.  Finally, it will be interesting to study the effects of
power corrections and hadronization on our results and to compare the
NNLO distributions of $O$ and EEC to data, thereby providing new
observables from which the value of the strong coupling $\alpha_s$ can
be extracted to NNLO accuracy.  

{\bf Acknowledgements:}
This research was supported by the Hungarian Scientific Research Fund grant
K-101482 and by the ERC Starting Grant ``MathAm''. 
AK gratefully acknowledges financial support from the
Post Doctoral Fellowship programme of the Hungarian Academy
of Sciences and the Research Funding Program ARISTEIA,
HOCTools (co-financed by the European Union (European
Social Fund ESF).


\end{document}